\documentclass[12pt]{revtex4}
  \usepackage{graphics, graphicx}
  \usepackage{color}

\begin{document}
 \def\ed{ ed. by}
 
 \title{Murray Gell-Mann:\\ A Short 
 Appreciation}
 
 \author{James B.~Hartle}

\email{hartle@physics.ucsb.edu}

\affiliation{Santa Fe Institute, Santa Fe, NM 87501}
\affiliation{Department of Physics, University of California,
 Santa Barbara, CA 93106-9530}

\date{\today }

\begin{abstract} On September 25, 2014 Murray Gell-Mann was presented with the Helmholz Medal of the Berlin Brandenburg Academy of Sciences and Humanities in a ceremony at the Santa Fe Institute. The author, among others, was asked to speak for fifteen minutes on Murray and his accomplishments. The following is an edited transcription of the author's speaking text.
\end{abstract}


\maketitle

 \large
 I am a student of Murray's and this year is the 50th anniversary of my Ph.D. under his supervision. In recent years he and I  have been working together to understand how our everyday classical world with all its complexity emerges from a simple quantum theory of the origin of the universe. 

I am not going to talk much about the universe or quantum mechanics. I am not even going to talk much about Murray's renowned  achievements in the physics of the elementary particles  --- strangeness, quarks, the eightfold way symmetries, the renormalization group, the standard model of the strong interactions (QCD) just to mention some. It is  for some of this work in particle physics that  he received the Nobel prize. It is the work for which he is perhaps best known. I'm not going to do that because others here  will tell you something about it and you can read about it in many places.  
But, more importantly,  to focus only on his work in particle physics would not do justice to the rest of his achievements. 

*His work on  the evolution of human languages.  Is there evidence in the many existing human languages that they all   had a common origin?   What does that tell you about human history? There is the Gell-Mann law for the evolution of the order of subject-object-verb for example.

*His seminal work on complex adaptive systems. We have no trouble exhibiting complex adaptive systems. The Santa Fe Institute is one. But what is there in common among the huge variety of  complex systems we find in the world?
What do we even mean by complexity? One answer is supplied  his beautiful work with Seth Lloyd on a general characterization of complexity. This is the kind of work that SFI was set up to do. It is encapsulated in the title of his book --- The Quark and the Jaguar. 

*His work on building a sustainable environment. A complex system that is central to our everyday life. 

*His work on the foundations of quantum mechanics and its application to cosmology. Quantum mechanics is the fundamental framework for the physics of the very small. It is  usually formulated for laboratory measurements carried out by observers. But in the early universe there were no observers. Are we to believe that quantum theory doesn't apply there?  No. In the very early universe large and small are one. At the big bang the  biggest possible system has the smallest possible size. It is obviously quantum mechanical. How then can quantum mechanics be formulated so it works at the big bang?  The decoherent histories quantum theory that Murray and I worked on is one answer. 

*And perhaps most importantly,  there is his work as a scientific citizen, especially with Presidents's Science Advisory Committee, the MacArthur Foundation, and his central role in setting up SFI. 

It is  an impressive and varied list. And I've often wondered what more he might have done in physics if he were more focussed on that.

What I hope to do  is to speculate (at great risk) on how these disparate areas  can be seen as part of a whole.
And to talk a little about what we can discern of Murray's approach to science from those connections and what we can learn about him from that. 

Erwin Schr\"odinger is often credited with the following injunction:

{\bf The task is not to see what no one else has seen, but to think what no one else has thought, about that which everyone else has seen.}
 
 That is Murray Gell-Mann. 
 
* To find the pattern that no one else had noticed was there. 

*To find the connections that no one else had sought. 

*To see the symmetry that is hidden or only manifested approximately. 

*To aim for the generality and the connections which no one else realized were possible.

*To have the courage to  give up on what was accepted as secure, general, and inescapable. 

* To know what is fundamental and what is excess baggage. 

* To  have the guts to guess the answer.  

It is  a common popular misconception that scientists working at the frontier are seeking to answer some well defined, fixed, famous great questions. ``What is the unified theory of all the fundamental interactions?''  ``What was the origin of the universe?''  But this is not how it seems in the  trenches.  

The frontier of theoretical physics is a noisy, chaotic place. At any one time, we have cherished old ideas, well confirmed in territory already mapped, that we are very reluctant to give up. We also have a whole variety of unconfirmed new ideas vying for the best route forward into new territory that {\it require} us to give up some of the old ideas. 

Rather the question is ``What {\it is} the  right question?''  Murray  always knows what the right question is. 
He  has a remarkable ability to see through all the clutter, to cut to the heart of matter, and to focus on the essentials. He also has the courage to discard the cherished old ideas that are an obstacle to progress. Later when looked at in the right way what he does seems inevitable. But that's the genius. 

From this perspective I think one can see some commonality in the subjects on which Murray has worked --- particle physics, human languages, complex adaptive systems, cosmology, and even the  environment.  They are all complicated with no, or non-obvious, or hidden regularities. But all  are systems for which Murray supplied some regularity.  

Even his avocations --- bird watching, collecting southwestern ceramics , wine tasting, etc. --- somehow fit in. 

The paradigmatic example is his work on the eightfold way symmetry of the strong interactions. At the time physicists were giving up on field theory so there was no real theory. But, Murray found the approximate symmetry relating the elementary particles {\it before} there was anything like a theory that had the symmetry. The symmetry predicted a previously unobserved particle, the particle was found, and a Nobel prize awarded.  

That led to quarks and eventually to string theory which is the focus of the efforts of many physicists today.

The power of  searching  for symmetries, connections and organizing principles that is characteristic of Murray's approach to science is well illustrated in this example. That  approach has had a 
lasting effect on how we do physics at the frontier.

{\bf ....to think what no one else has thought, about that which everyone else has seen.}

 \begin{figure}[t]
\includegraphics[width=6in]{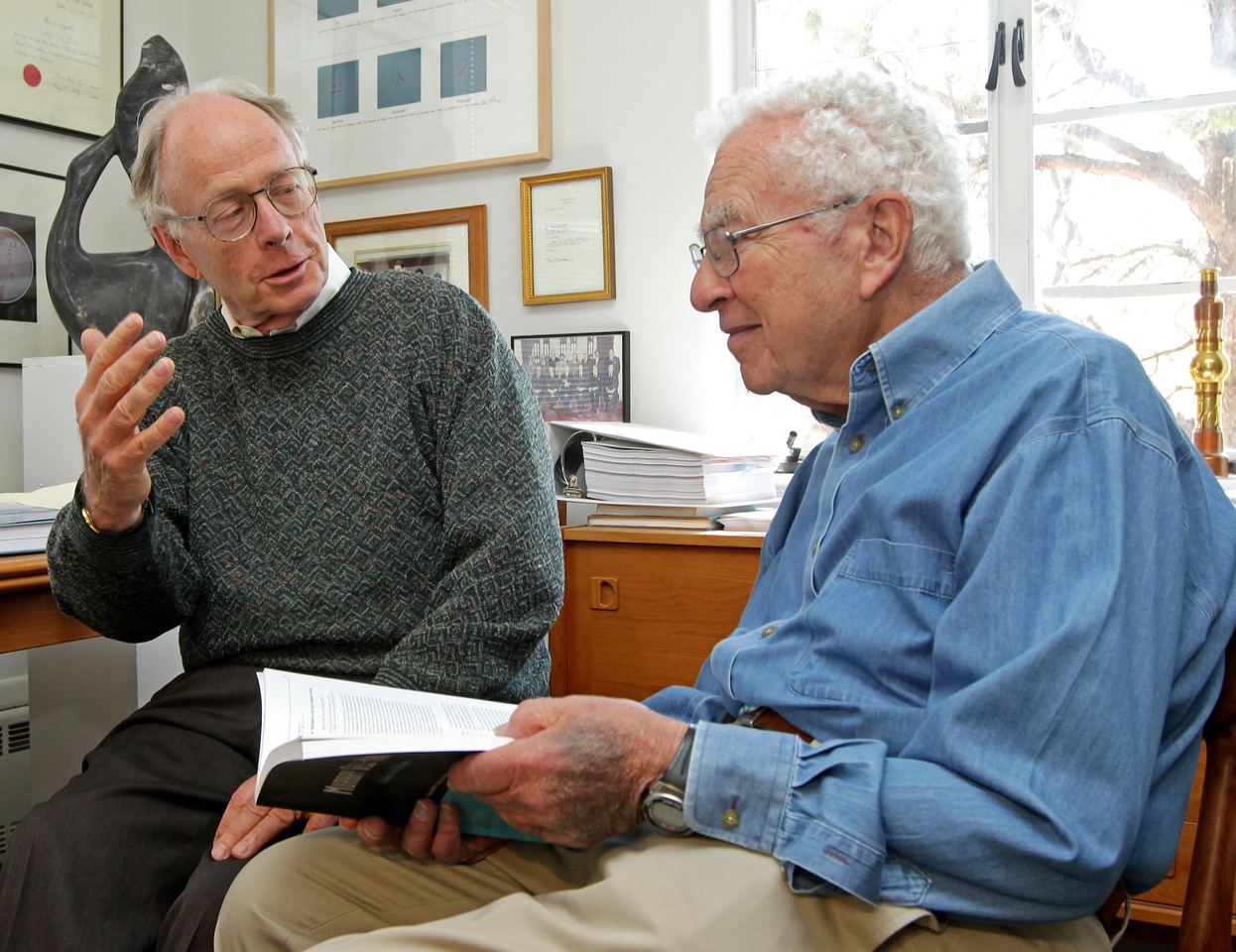}
\caption{The author and Murray 
working in Murray's office at SFI  in 2015. \\ Photo courtesy SFI.jpeg}
\label{L1042-c}
\end{figure}

Working with Murray is a unique kind of experience, at least in my life.  To work with him it  helps first of all not to have too fragile an ego. Murray was very accessible to his students. But I remember one occasion when I went to his office and said ``Professor Gell-Mann, I have a question.'' He replied somewhat irritably  ``Questions, questions --- you are always coming in here with questions! Why don't you try bringing in some answers sometime!''  Caltech wasn't called the Marine Corps of graduate schools for nothing. But despite the obvious differences in intellect and insight Murray somehow makes you feel an equal partner when you are working with him.
 
Working with Murray you have to accept certain rigorous  standards. There is typically a list of things Murray is sure are  connected and we go over and over them again and again,  posing them, attacking them,  calculating them until a unified, new, surprising perspective emerges.  Never stopping short, never settling for a picture that is incomplete,  it can take years.

Working with Murray is challenging, but it is also fun!   To a good approximation Murray knows everything. The conversation is not only about physics, but about languages, the history of science, the history of the Ottoman empire, politics, the lives of colleagues,  the environment, philosophy, etc. You enter a world in which everything is connected on a much bigger scale and in a much deeper way  than in your narrow speciality. You make personal progress not only in science but in almost everything else. It  is a true adventure of the mind. And the same thing could be said about Murray's life. 

In the future, as physicists, Murray will always be in our minds, because he is a great scientist, because he created much of what we have to think about, and  because he set a certain style on how to do science. But for those of us who have had the privilege of working with him he will also  be --- always in our hearts. Thank you.

\appendix

\centerline{\bf Papers with Murray Gell-Mann}

\begin{itemize}

\item{} {\it Quantum Mechanics in the Light of Quantum Cosmology} ,  in {\sl
Complexity, Entropy, and the Physics of Information, Santa Fe Institute
Studies in the Sciences of Complexity VIII}, ed. by  W.H. Zurek, (Addison-Wesley,
Reading, MA  1990) and in  {\sl Proceedings of the 3rd International Symposium on the Foundations of Quantum Mechanics in the Light of New Technology}, ed. by S. Kobayashi, H. Ezawa, Y. Murayama, and S. Nomura, (Physical Society of Japan, Tokyo, 1990), arXiv:1803.04605.

\item{}  {\it Alternative Decohering Histories in Quantum Mechanics} 
in the {\sl Proceedings of
the 25th International Conference on High Energy Physics, Singapore,
August,
2-8, 1990}, 
\ed K.K.~Phua and Y.~Yamaguchi (South East Asia Theoretical 
Physics Association
and Physical Society of Japan distributed by World Scientific,
Singapore,1990).

\item{} {\it Time Symmetry and Asymmetry in Quantum Mechanics and Quantum
Cosmology},
 in the {\sl Proceedings of the 1st International
Sakharov Conference on Physics}, Moscow, USSR, May 27--31, 1991, \ed
L.V.~Keldysh and V.Ya.~Fainberg, Nova
Science Publishers, New York (1992); and  in
{\sl Physical Origins of Time Asymmetry: Proceedings of the NATO
Workshop},
Mazagon, Spain, September 30--October 4,
1991,
\ed J.~Halliwell, J.~Perez-Mercader, and W.~Zurek, Cambridge
University Press, Cambridge (1994) pp.~311-345; arXiv:gr-qc/9304023.

\item{} {\it Classical Equations for Quantum Systems},  {\sl Phys. Rev., D} {\bf 47}, 3345--3382, 1993; arXiv:gr-qc/9210010.

\item{}{\it Equivalent Sets of Histories and Multiple Quasiclassical
Domains}; arXiv:gr-qc/9404013.

\item{} {\it Strong Decoherence} in the
{\sl Proceedings of the 4th (1994)
Drexel Conference on Quantum Non-Integrability:
Quantum-Classical Correspondence}, \ed D.-H.~Feng and B.-L.~Hu,
International Press  \\ of Boston/Hong Kong (1998);
arXiv:gr-qc/9509054.

\item{}  {\it Quasiclassical Coarse Graining and Thermodynamic Entropy} , {\sl Phys. Rev. A}, {\bf 76}, 022104 (2007) , arXiv:quant-ph/0609190.

\item{}  {\it Decoherent Histories Quantum Mechanics with One `Real' Fine-Grained History}, {\sl Phys. Rev. A}, {\bf 85}, 062120 (2012); arXiv:1106.0767.

\item{}  {\it Adaptive Coarse Graining, Environments, Strong Decoherence, and  Quasiclassical Realms}, {\sl Phys. Rev. A}, {\bf 89}, 052125 (2014); \\ arXiv:1312.7454.

\end{itemize}

 \end{document}